\documentclass[useAMS,usenatbib]{mn2e}

\usepackage[T1]{fontenc}
 \usepackage{pslatex}
\usepackage{amsmath}
\usepackage{amsfonts}
\usepackage{color}
\usepackage{url}
\usepackage{bm}
\usepackage{graphicx}
\usepackage{subfigure}
\usepackage{amssymb}
\usepackage{soul}
\usepackage{booktabs}
\usepackage{array}
\usepackage[utf8]{inputenc}
\inputencoding{latin1}
\inputencoding{utf8}

{\newif\ifnotend
\notendtrue
\def\veclist{ABCDEFGHIJKLMNOPQRSTUVWXYZabcdefghijklmnopqrstuvwxyz.}
\def\top#1#2.{#1}
\def\tail#1#2.{#2.}
\loop\expandafter\xdef\csname v\expandafter\top\veclist\endcsname%
{{\noexpand\bf\expandafter\top\veclist}}
\edef\veclist{\expandafter\tail\veclist}
\if\veclist.\notendfalse\fi\ifnotend\repeat}
\def\e{{\rm e}}

\def\E{{\cal E}}

\mathchardef\mhyphen="2D

\title[On the properties of BL Lac jets]{On the magnetisation and the radiative efficiency of BL Lac jets}
\author[Sobacchi \& Lyubarsky]{E. Sobacchi$^{1,2}$\thanks{E-mail: sobacchi@post.bgu.ac.il} \& Y. E. Lyubarsky$^1$\\
$^1$ Physics Department, Ben-Gurion University, P.O.B. 653, Beer-Sheva 84105, Israel \\
$^2$ Department of Natural Sciences, The Open University of Israel, 1 University Road, P.O.B. 808, Raanana 4353701, Israel
}
\begin{document}

\date{}

\def\p{\partial}
\def\E{\textbf{E}}
\def\B{\textbf{B}}
\def\v{\textbf{v}}
\def\j{\textbf{j}}
\def\s{\textbf{s}}
\def\e{\textbf{e}}

\newcommand{\di}{\mathrm{d}}
\newcommand{\bfx}{\mathbf{x}}
\newcommand{\bfe}{\mathbf{e}}
\newcommand{\vlos}{\mathrm{v}_{\rm los}}
\newcommand{\Tspin}{T_{\rm s}}
\newcommand{\Tb}{T_{\rm b}}
\newcommand{\degree}{\ensuremath{^\circ}}
\newcommand{\Th}{T_{\rm h}}
\newcommand{\Tc}{T_{\rm c}}
\newcommand{\bfr}{\mathbf{r}}
\newcommand{\bfv}{\mathbf{v}}
\newcommand{\bfu}{\mathbf{u}}
\newcommand{\pc}{\,{\rm pc}}
\newcommand{\kpc}{\,{\rm kpc}}
\newcommand{\Myr}{\,{\rm Myr}}
\newcommand{\Gyr}{\,{\rm Gyr}}
\newcommand{\kms}{\,{\rm km\, s^{-1}}}
\newcommand{\de}[2]{\frac{\partial #1}{\partial {#2}}}
\newcommand{\cs}{c_{\rm s}}
\newcommand{\rb}{r_{\rm b}}
\newcommand{\rqu}{r_{\rm q}}
\newcommand{\bfOmega}{\pmb{\Omega}}
\newcommand{\bfOmegap}{\pmb{\Omega}_{\rm p}}
\newcommand{\bfXi}{\boldsymbol{\Xi}}

\maketitle

\begin{abstract}
Theoretical modelling and observations of AGN jets suggest that the non-thermal electrons emitting the observed radiation should (i) carry an amount of energy comparable to the magnetic fields ($U_e\sim U_B$), which is likely the case if the magnetic fields play a dynamically important role in the jet's acceleration process; (ii) cool efficiently in a dynamical time ($t_{\rm cool}\lesssim t_{\rm dyn}$), which is suggested by the fact that a large fraction of the jet's kinetic energy is promptly converted into radiation. These expectations are at odds with the results of the simplest one-zone Self-Synchro-Compton (SSC) model for the Spectral Energy Distribution (SED) of BL Lacs. Indeed, the model predicts $U_e\gg U_B$ and $t_{\rm cool}\gg t_{\rm dyn}$ for most of the objects. Here we closely investigate one of the key assumptions of this model, namely that the momentum distribution of the non-thermal electrons is isotropic. We find that this assumption may be an oversimplification. If the magnetic energy is dissipated via a turbulent MHD cascade, the highest energy electrons may instead retain a small pitch angle. Since the synchrotron emissivity is suppressed when the pitch angle is small, this effect may importantly affect the modelling of the SED. As an illustrative example, we present an anisotropic model for the electron momentum distribution such that $U_e\sim U_B$ and $t_{\rm cool}\lesssim t_{\rm dyn}$ at the same time. Our model manages to simultaneously solve the two problems with one only more free parameter with respect to the usual isotropic one-zone SSC model.
\end{abstract}



\section{Introduction}
\label{sec:introduction}

The super massive black holes residing in the centre of galaxies are able to launch jets that reach relativistic velocities. These jets produce remarkably non-thermal spectra. Observations suggest that a significant fraction (typically $\sim 15\%$) of the jet's total energy is promptly radiated \citep[e.g.][]{Nemmen2012}. In order to achieve such a high radiative efficiency, the energy dissipation process must involve the efficient acceleration and the subsequent fast cooling of a population of non-thermal particles.

AGN jets are thought to be powered by the rotational energy of the super massive black hole, which is channeled into the outflowing plasma via electromagnetic stresses \citep[e.g.][]{Blandford1976, Lovelace1976, BlandfordZnajek1977}. In this scenario the energy budget of the jet is initially dominated by the Poynting flux, which is gradually converted into the plasma kinetic energy while the flow is accelerated. Theoretical investigation of the energy dissipation process suggests that in this regime the emitting electrons and the electromagnetic fields may carry comparable amounts of energy \citep[see for example][]{Sironi2015}.

Among blazars (i.e. AGN with the jet pointing towards the observer), BL Lacs are the ideal laboratory to test our theoretical understanding of the physics of relativistic jets. These objects are particularly attractive due to their simplicity, since they do not show any emission from the black hole accretion disc, as sometimes is the case for Flat Spectrum Radio Quasars (FSRQ). The Spectral Energy Distribution (SED) of BL Lacs can therefore be simply interpreted as due to Self-Synchro-Compton emission from a population of non-thermal electrons \citep[e.g.][]{Tavecchio2010}.

In order to fit the SED, one has to assume a broken power law for the energy distribution of the electrons. One needs just to specify (i) the number density of the non-thermal electrons; (ii) the Lorentz factor $\gamma_{\rm b}$ of the electrons at the break; (iii) the size $R$, (iv) the magnetic field $B$, and (v) the Doppler factor $\delta$ of the dissipation region. As discussed by \citet{Tavecchio1998}, all the parameters of the model are univocally determined by the observed SED.

Despite the apparent simplicity of this model, it is challenging to give a convincing physical interpretation of the results. The reason for this is twofold \citep[see for example][]{TavecchioGhisellini2016}: (i) the model predicts that the magnetic fields carry just a tiny fraction of the electrons energy, contrary to the theoretical expectation that the two energies are in an approximate equipartition; (ii) the cooling time of the electrons at the break is much longer than the dynamical time, which imply that the jet is extremely radiatively inefficient and raises the further problem to explain the origin of the break itself. Note that a low radiative efficiency would be difficult to reconcile with the observations of \citet{Nemmen2012}, who instead found that a significant fraction of the jet's energy is promptly radiated.

It has been suggested that relaxing the assumption of a one-zone emission model may help to solve this controversy \citep[see for example][]{TavecchioGhisellini2016}. Though a two-zone emission model can likely provide $U_e\sim U_B$ and $t_{\rm cool}\lesssim t_{\rm dyn}$ for any individual object, we argue that extending this interpretation to the entire BL Lac sample may require some fine tuning. Indeed, since in BL Lacs the luminosity of the synchrotron peak is typically comparable to the luminosity of the IC peak of the SED \citep[e.g.][]{Tavecchio2010}, one would expect the magnetic energy density $U_B$ to be comparable to the radiation energy density $U_\gamma$. In any model with $U_e\sim U_B$, this immediately implies that $U_e\sim U_\gamma$. In a one-zone model, $U_e\sim U_\gamma$ is naturally achieved if the electrons efficiently radiate their energy in a dynamical time. However, if $U_\gamma$ is due to an external radiation field, as is the case in two-zone emission models, it seems to us that there is not any good a priori reason why $U_e\sim U_\gamma$.

The tension mentioned above motivates a closer examination of the one-zone SSC emission model that is usually adopted to interpret the SED of BL Lacs. Here we challenge one of the key assumptions of this paradigm, namely that the momentum distribution of the non-thermal electrons is isotropic. We show that, if the magnetic energy is dissipated via a turbulent MHD cascade, the highest energy electrons may retain a small pitch angle, which suppresses their synchrotron emission. Taking this effect into account may importantly affect the modelling of the BL Lac SED. Indeed, as an illustrative example we present an anisotropic model for the electron momentum distribution such that (i) the non-thermal electrons and the magnetic fields carry comparable amounts of energy, and (ii) the electrons at the break efficiently cool in a dynamical time. Our model has just one more free parameter than the standard isotropic model.

The paper is organised as follows. In Section \ref{sec:motivation} we discuss the reason why the highest energy electrons may retain a small pitch angle. In Section \ref{sec:model} we describe the predictions of our anisotropic model for the electron momentum distribution. In Section \ref{sec:comparison} we compare these predictions with a model that instead assumes an isotropic momentum distribution. In Section \ref{sec:results} we present our main results. Finally, in Section \ref{sec:conclusions} we summarise our conclusions. Throughout this paper we always work in the frame of the source, or equivalently we assume that the source is at redshift $z=0$.

\section{Motivation for an anisotropic model}
\label{sec:motivation}

A crucial point in the hydromagnetic jet launching paradigm is understanding how the magnetic energy is dissipated. Since the physical scale of the jet typically exceeds the Larmor radius of the non-thermal particles by many orders of magnitude, it is natural to assume that the energy is brought down to the dissipation scale by a turbulent MHD cascade. In the following we assume the cascade to be injected at the outer scale $R$ of the dissipation region.\footnote{Such a turbulent cascade may be triggered by MHD instabilities in a Poynting-dominated jet. Recent PIC simulations have shown that highly tangled magnetic fields may be formed in the kink-unstable region of the jet, resulting in the dissipation of the magnetic energy and the rapid acceleration of a population of non-thermal particles \citep[e.g.][]{Alves2018, Nalewajko2018}.} In a relativistic, optically thin plasma the photon viscosity is unable to damp the cascade, which should then proceed unimpeded down to microscopic scales \citep[see for example][]{Zrake2018}.

\subsection{Dissipation of MHD turbulence leads to longitudinal particle heating}

The most important property of MHD turbulence is its strong anisotropy, with the turbulent eddies becoming strongly elongated in the direction of the background magnetic field at small scales. As first proposed by \citet{GoldreichSridhar1995}, in the course of the turbulent cascade the ratio of the longitudinal scale of the eddies, $\lambda_\parallel$, to the Alfv\'{e}n velocity remains equal to the ratio of the perpendicular scale, $\lambda_\perp$, to the turbulent velocity (this condition is known as ``critical balance''). From this condition, one finds that $\lambda_\perp$ and $\lambda_\parallel$ are related by
\begin{equation}
\lambda_\perp/\lambda_\parallel \sim \sqrt{\lambda_\parallel/R}\;,
\end{equation}
where $R$ is the outer scale, while the cascade is described by a Kolmogorov spectrum in the perpendicular direction.

\citet{ThompsonBlaes1998} extended the theory to the extreme relativistic regime, when the plasma inertia is negligible (force-free MHD). They argued that an anisotropic cascade is formed also in this case, and that the dissipation occurs at the scale of the current starvation, i.e. when there are not enough charge carriers in the plasma to maintain the currents associated with the Alfv\'{e}n waves. As pointed out by \citet{Thompson2006}, in this case the dissipation of relativistic MHD turbulence heats the particle in the longitudinal direction, and a particle distribution that is strongly elongated in the direction of the background magnetic field might therefore be expected. \citet{Thompson2006, ThompsonGill2014, GillThompson2014} suggested that the rapid variability of the GRB prompt emission may be attributed to this anisotropy.

The statement that in collisionless plasmas the anisotropic MHD turbulence decays by heating/accelerating particles along the background magnetic field is general. Indeed, the dissipation occurs at the wave-particle resonances
\begin{equation}
\label{eq:res}
\omega - {\bf k}\cdot{\bf v} = n\Omega_{\rm L} \;,
\end{equation}
where $\omega$ and ${\bf k}$ are the frequency and the wavenumber, ${\bf v}$ is the particle velocity, $\Omega_{\rm L}\equiv eB/\gamma mc$ is the particle relativistic Larmor frequency, and $n$ is an integer. The cyclotron resonance condition, $n\neq 0$, is satisfied when the longitudinal scale of the wave packet, $\lambda_\parallel$, is of the order of the particle Larmor radius, $r_{\rm L}\equiv c/\Omega_{\rm L}$. Since typically $r_{\rm L}\ll R$, due to the strong anisotropy of MHD turbulence a particle crosses many wave packets during one Larmor orbit and the energy gain averages out. Hence, wave-particle interactions mediated by the cyclotron resonance can be neglected. It was first noticed by \citet{Gruzinov1998, Quataert1998, QuataertGruzinov1999} that in this case the dissipation occurs at the Landau resonance, $n=0$. Since the two physical mechanisms of wave-particle interaction at the $n=0$ resonance are due to (i) the longitudinal electric field of the wave and (ii) the interaction between the effective particle's magnetic moment and the longitudinal magnetic perturbation, one is led to the conclusion that the turbulent energy is primarily dissipated onto the longitudinal particle motion.

It has also been found that the turbulent fluctuations tend to align with one another forming small scale current sheets \citep[e.g.][]{Boldyrev2006, BeresnyakLazarian2006, Mason2006}, which could be disrupted via magnetic reconnection thus providing an additional dissipation mechanism \citep[e.g.][]{BoldyrevLoureiro2017, Mallet2017a, Mallet2017b, LoureiroBoldyrev2017}. Note that the background magnetic field, which is much larger than the reconnecting field and lies in the same plane of the current sheet, plays the role of a guide field. Since the magnetic energy is transferred to the plasma particles at the Landau resonance between the particles and the tearing mode that disrupts the current sheet, also in this case one would expect the particles to be heated in the longitudinal direction.

Even if the perpendicular heating is negligible, in a weakly magnetised plasma the fire-hose instability quickly erases any momentum anisotropy \citep[e.g.][]{Parker1958, Lerche1966}. However, since the fire-hose instability develops once $P_\parallel-P_\perp>B^2/4\pi$ (where $P_\parallel$ and $P_\perp$ are the parallel and perpendicular pressure components respectively), one immediately sees that this instability is not effective if the magnetic to plasma energy ratio exceeds $1/2$. Moreover, a significant angular spread, say $\langle\sin^2\theta\rangle\sim 1/2$ where $\theta$ is the particle pitch angle, is expected only when the energy ratio drops to very small values. In the following we argue that in the highly magnetised regime (i) some plasma instability may still be able to make the electron momentum distribution isotropic; (ii) such an instability may be ineffective for the most energetic electrons, which concludes our argument.

\subsection{Momentum isotropisation in a highly magnetised plasma by the resonance instability of Alfv\'{e}n waves}

If the energy of the system was initially stored in the magnetic field, most of the energy release occurs around the equipartition stage when the fire-hose instability does not work. Another isotropisation mechanism is the cyclotron instability that develops at the anomalous cyclotron resonance ($n=-1$ in Eq. \ref{eq:res}). In this case, particles with super-Alfv\'{e}n velocities excite Alfv\'{e}n waves and at the same time their pitch angle increases, the energy being taken from the longitudinal motion. This process is analogous to the classic cosmic-ray scattering by Alfv\'{e}n waves \citep[e.g.][]{Lerche1967, KulsrudPearce1968}.

In the following we assume the momenta of all the particles to be initially directed along the magnetic field. To fix ideas, let the background magnetic field $B$ be directed in the positive $z$ direction. The Alfv\'{e}n wave velocity $v_{\rm A}$ is mildly relativistic at the equipartition stage. In order to determine the stability properties of the plasma, we adapt an argument due to \citet{Kulsrud2005} to the case we are interested in.

\subsubsection{Electron-positron plasma}

We first study the case of an electron-positron plasma. Let us consider a right circularly polarised wave packet with wavelength $\lambda_\parallel$ propagating in the positive $z$ direction. Such a wave is emitted by the positrons at the $n=-1$ resonance, which propagate in the positive $z$ direction, and is absorbed by the electrons at the $n=1$ resonance, which propagate in the negative $z$ direction. Using $n=-1$ in Eq. \eqref{eq:res}, we may calculate the Lorentz factor of the resonant positrons as
\begin{equation}
\gamma_{e^+ {\rm ,res}}\sim \frac{1}{1-\beta_{\rm A}}\frac{eB\lambda_\parallel}{m_e c^2}\;,
\end{equation}
where $\beta_{\rm A}\equiv v_{\rm A}/c$. In a similar way, the Lorentz factor of the resonant electrons is obtained substituting $n=1$ into Eq. \eqref{eq:res}, which gives
\begin{equation}
\gamma_{e^- {\rm ,res}}\sim \frac{1}{1+\beta_{\rm A}}\frac{eB\lambda_\parallel}{m_e c^2}\;.
\end{equation}
Assuming that the pairs are distributed according to a power law with energy index $\sim -2$, which is ultimately motivated by the observed SED \citep[e.g.][]{Tavecchio2010}, one finds the number density of resonant positrons,
\begin{equation}
\label{eq:nres_0}
n_{e^+{\rm,res}}\sim \frac{n_e}{2\gamma_{e^+ {\rm ,res}}} \sim n_e \frac{1-\beta_{\rm A}}{2} \frac{m_e c^2}{eB\lambda_\parallel}\;,
\end{equation}
and the number density of resonant electrons,
\begin{equation}
\label{eq:nres_01}
n_{e^-{\rm,res}}\sim \frac{n_e}{2\gamma_{e^- {\rm ,res}}} \sim n_e \frac{1+\beta_{\rm A}}{2} \frac{m_e c^2}{eB\lambda_\parallel}\;,
\end{equation}
where $n_e\sim n_{e^+}+n_{e^-}$ is the total number density of the pairs.

The momentum of the resonant positrons is $p_{e^+{\rm, res}}\sim \gamma_{e^+ {\rm ,res}}m_ec$, while the momentum of the resonant electrons is $p_{e^-{\rm, res}}\sim \gamma_{e^- {\rm ,res}}m_ec$. Let $\delta B$ be the amplitude of the wave packet. Since the pitch angle of the positrons diffuses on a time scale $t_{e^+{\rm,diff}}\sim \Omega_{\rm L}^{-1}\left(\delta B/B\right)^{-2}\sim \left(\gamma_{e^+ {\rm ,res}}m_e c/eB\right)\left(\delta B/B\right)^{-2}$ and the pitch angle of the electrons diffuses on a time scale $t_{e^-{\rm,diff}}\sim \left(\gamma_{e^- {\rm ,res}}m_e c/eB\right)\left(\delta B/B\right)^{-2}$, one finds
\begin{equation}
\label{eq:emission_p}
\frac{\Delta p_{e^+{\rm ,res}}}{\Delta V \Delta t} \sim \frac{n_{e^+{\rm ,res}}p_{e^+{\rm, res}}}{t_{e^+{\rm,diff}}} \sim eB n_{e^+{\rm ,res}} \left(\frac{\delta B}{B}\right)^2\;,
\end{equation}
and
\begin{equation}
\label{eq:absorption_p}
\frac{\Delta p_{e^-{\rm ,res}}}{\Delta V \Delta t} \sim \frac{n_{e^-{\rm ,res}}p_{e^-{\rm, res}}}{t_{e^-{\rm,diff}}} \sim eB n_{e^-{\rm ,res}} \left(\frac{\delta B}{B}\right)^2\;.
\end{equation}
These are respectively the momentum density gained (lost) by the wave per unit time due to the resonant interaction with the positrons (electrons). Since $n_{e^+{\rm ,res}}<n_{e^-{\rm ,res}}$, the emission term \eqref{eq:emission_p} is smaller than the absorption term \eqref{eq:absorption_p} and the wave is damped.

Hence, in an electron-positron plasma the instability does not develop and we expect the particle distribution to remain strongly elongated in the direction of the background magnetic field. The synchrotron emissivity depends on the magnetic field through the combination $B\sin\theta$, where $\theta$ is the pitch angle \citep[e.g.][]{RybickiLightman1979}. Hence, we see that if $\theta\ll 1$ the magnetic field in the dissipation region might be significantly stronger than what it is inferred assuming an isotropic momentum distribution for the non-thermal electrons. Note, however, that the cooling time remains the same even in the limit $\theta\ll 1$.

\subsubsection{Electron-positron-ion plasma}

The presence of an even small (in terms of number density) ion component may completely change the results obtained for an electron-positron plasma. It is important to realise that the amplitude of the right circularly polarised wave considered in the previous section grows due to the resonant interaction with the protons moving in the positive $z$ direction. The fundamental difference with respect to the pair plasma is that there are not negatively charged ions, which would be the analogous of the electron component, that damp the wave. Hence, the amplitude of the wave grows if the number of resonant protons and positrons, which emit the wave, exceeds the number of resonant electrons, which absorb the wave.

In the following we consider the case when the pairs dominate the number density ($n_e\gg n_p$), but the protons dominate the rest mass density of the jet ($n_p m_p\gg n_e m_e$), which is motivated by a number of independent arguments in the literature \citep[e.g.][]{SikoraMadejski2000, GhiselliniTavecchio2010}. We assume that (i) the pairs are distributed according to a power law with energy index $\sim -2$, and (ii) the proton energy distribution is steeper than the electron one, as discussed in more detail below, in which case most of the proton's energy is carried by mildly relativistic particles. Hence, the proton to electron energy density is $U_p/U_e \sim n_pm_p/n_e m_e \log\left(\gamma_{\rm b}\right)$. Using a typical break Lorentz factor $\gamma_{\rm b}\sim 10^4-10^6$ for the non-thermal pairs \citep[e.g.][]{Tavecchio2010}, one finds $U_p/U_e\sim \left(100-200\right)\times n_p/n_e$. Hence, if $n_e\sim \left(10-100\right)\times n_p$ \citep[e.g.][]{SikoraMadejski2000}, the energy carried by the protons does not typically exceed that carried by the pairs by a large factor.

Using the same argument as to derive Eq. \eqref{eq:emission_p}-\eqref{eq:absorption_p}, one can calculate the momentum density gained by the wave per unit time due to the interaction with the resonant ($n=-1$) protons, which gives
\begin{equation}
\label{eq:pwave1}
\frac{\Delta p_{p{\rm ,res}}}{\Delta V\Delta t} \sim e B n_{p{\rm,res}} \left(\frac{\delta B}{B}\right)^2\;.
\end{equation}
Combining Eqs. \eqref{eq:emission_p}-\eqref{eq:pwave1}, one sees that the emission is larger than the absorption, and hence the wave grows, if
\begin{equation}
\label{eq:inst}
n_{p{\rm,res}} + n_{e^+{\rm,res}} \gtrsim n_{e^-{\rm,res}}\;.
\end{equation}
This is the condition for a particle distribution with all the momenta directed along the background magnetic field to be unstable. If the instability develops, the electrons are isotropised by the absorption of the resonant waves, while the positrons and the protons are isotropised by the emission. In Appendix \ref{sec:appendix} we calculate the growth rate of the instability, showing that it is fast enough (with respect, for example, to the dynamical time) for the instability to be indeed effective once the condition \eqref{eq:inst} is satisfied.

In order to make further progress we need to make some assumptions on the proton energy distribution, which determines $n_{p{\rm ,res}}$. First of all, note that if the wavelength is shorter than the proton non-relativistic Larmor radius ($\lambda_\parallel\lesssim m_pc^2/eB$), the number density of the resonant protons equals the total number density of the protons, namely $n_{p{\rm,res}}\sim n_p$. For longer wavelengths, $n_{p{\rm,res}}$ depends on the details of the heating process.

In the following we assume that the protons are distributed according to a power law with energy index $-s$. We take $s>2$, namely we assume that the proton distribution is steeper than the pair distribution. This choice is motivated by the fact that, as a result of the dissipation of non-relativistic MHD turbulence, the proton to electron heating ratio is a decreasing function of the magnetisation, and is already smaller than unity when the thermal and the magnetic energy are in equipartition \citep[e.g.][]{QuataertGruzinov1999, Howes2010}. However, one should realise that the extrapolation of these results to the relativistic regime is far from obvious, and would require further investigation.

Since the Lorentz factor of the resonant protons is $\gamma_{p{\rm ,res}}\sim eB\lambda_\parallel/m_pc^2$, one finds
\begin{equation}
\label{eq:nres_1}
n_{p{\rm,res}}\sim n_p \times
\begin{cases}
1 & {\rm if}  \quad \lambda_\parallel\lesssim m_pc^2/eB\\
\left(\frac{m_p c^2}{eB\lambda_\parallel}\right)^{s-1} & {\rm if} \quad \lambda_\parallel\gtrsim m_pc^2/eB \;.
\end{cases}
\end{equation}
The important point is that the number of the resonant protons can exceed the number of resonant pairs ($n_{p{\rm,res}}\gtrsim n_{e{\rm,res}}$) even when the pairs dominate the total number density ($n_p\ll n_e$). The reason for this is the large proton to electron mass ratio, $m_p/m_e\gg 1$, which implies that the Lorentz factor of the protons resonating with a given $\lambda_\parallel$ is significantly smaller than the Lorentz factor of the pairs resonating with the same wave.

Using Eqs. \eqref{eq:nres_0} and \eqref{eq:nres_01} for $n_{e^+{\rm ,res}}$ and $n_{e^-{\rm ,res}}$, and Eq. \eqref{eq:nres_1} for $n_{p{\rm ,res}}$, we see that the condition \eqref{eq:inst} is equivalent to
\begin{equation}
\label{eq:lambda_res_1}
\frac{n_e m_e}{n_p m_p}\frac{m_pc^2}{eB} \lesssim\lambda_\parallel\lesssim \left(\frac{n_p m_p}{n_e m_e}\right)^{\frac{1}{s-2}}\frac{m_pc^2}{eB}\;,
\end{equation}
where we have used the fact that $\beta_{\rm A}$ is of order unity. The Lorentz factor of the pairs resonating with the largest unstable $\lambda_\parallel$ can be found from $\gamma_{\rm iso}m_e c^2/eB\sim \lambda_\parallel$, which finally gives
\begin{equation}
\label{eq:gamma_iso}
\boxed{\gamma_{\rm iso}\sim \left(\frac{n_p m_p}{n_e m_e}\right)^{\frac{1}{s-2}}\frac{m_p}{m_e} }\;.
\end{equation}
The pairs with $\gamma\lesssim\gamma_{\rm iso}$ are isotropised due to the resonant interaction with the waves, while those with $\gamma\gtrsim\gamma_{\rm iso}$ retain a small pitch angle and thus do not radiate by synchrotron.

Since $\gamma_{\rm iso}\gtrsim m_p/m_e \sim 2\times 10^3$, it is possible that $\gamma_{\rm iso}\lesssim\gamma_{\rm b}$ in a significant fraction of BL Lacs, where the break Lorentz factor can be as large as $10^5-10^6$ \citep[e.g.][]{Tavecchio2010}. As we discuss in the next section, this fact is important for the modelling of the BL Lac SED. Finally, note that the exact value of $\gamma_{\rm iso}$ depends on the poorly known details of the particle heating in relativistic MHD turbulence, which determine $s$, and on the composition of the jet.

\section{Predictions of the model}
\label{sec:model}

\begin{figure}{\vspace{3mm}} 
\centering
\includegraphics[width=0.49\textwidth]{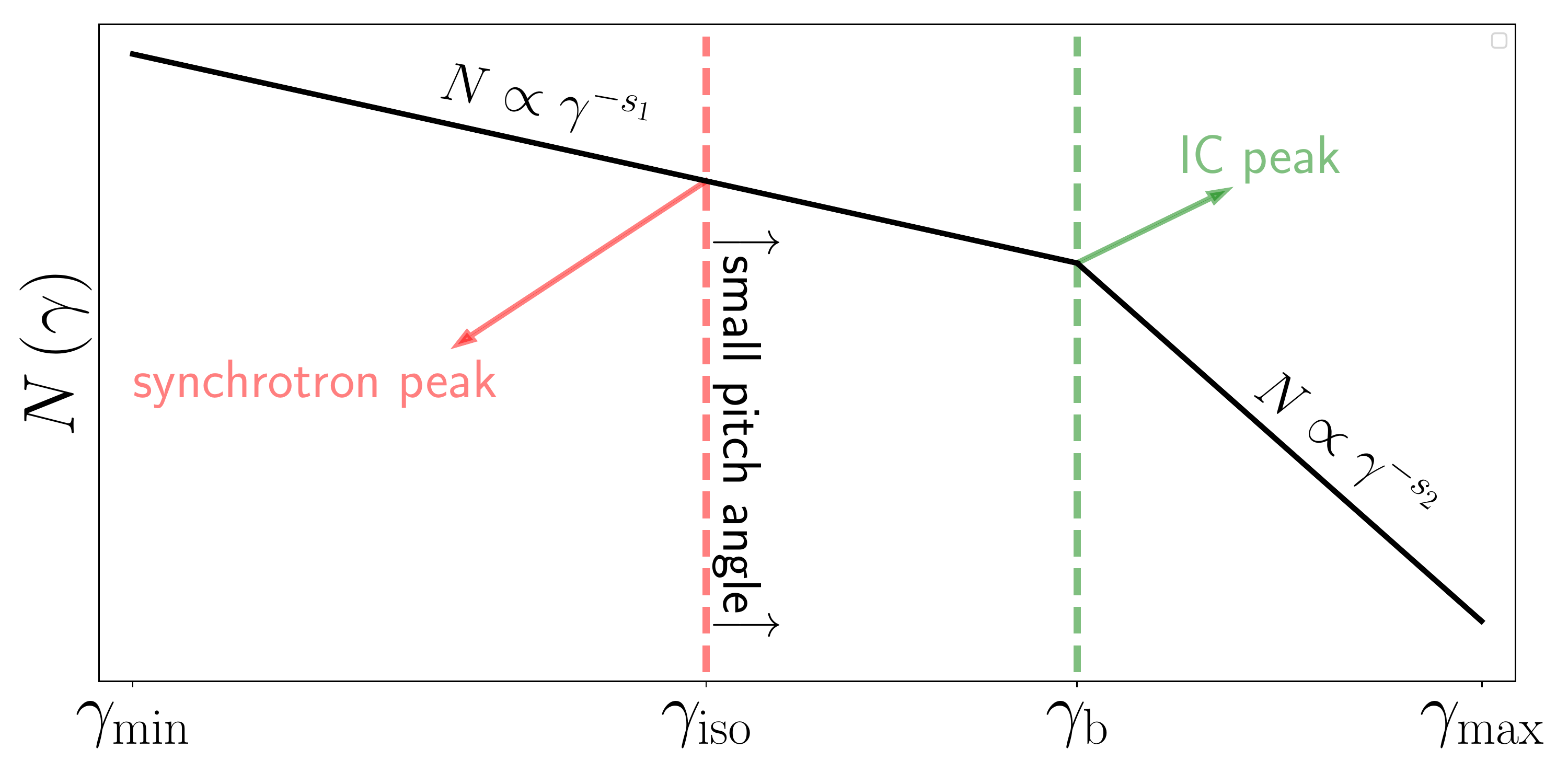}
\caption{Main features of our model. The energy distribution of the electrons in BL Lac jets is described by a broken power law (see Eq. \ref{eq:N}) extending from $\gamma_{\rm min}$ to $\gamma_{\rm max}$. The electrons at the break ($\gamma\sim\gamma_{\rm b}$) produce the Inverse Compton peak of the SED. The distribution becomes strongly elongated in the direction of the magnetic field for $\gamma\gtrsim\gamma_{\rm iso}$. Since the synchrotron emission by the electrons with $\gamma\gtrsim\gamma_{\rm iso}$ is suppressed due to their small pitch angles, when $\gamma_{\rm iso}\lesssim\gamma_{\rm b}$ the synchrotron peak of the SED is produced by the electrons with $\gamma\sim\gamma_{\rm iso}$.}
\label{fig:model}
\end{figure}

The main features of our model are sketched in Figure \ref{fig:model}. In the following all the physical quantities are defined in the frame of the dissipation region. Following \citet{Tavecchio1998}, we assume that the energy distribution of the non-thermal electrons is described by
\begin{equation}
\label{eq:N}
N\left(\gamma\right) = 
\begin{cases}
K \gamma^{-s_1} & {\rm if}\quad \gamma_{\rm min}<\gamma<\gamma_{\rm b}\\
K \gamma_{\rm b}^{s_2-s_1} \gamma^{-s_2} & {\rm if}\quad \gamma_{\rm b}<\gamma<\gamma_{\rm max}
\end{cases}
\end{equation}
where the scaling constant $K$ has units of cm$^{-3}$. The spectral indices $s_1$ and $s_2$ can be determined directly from the SED; one typically finds $s_1= 1.8-2.2$ and $s_2 = 3.5-5$ \citep{Tavecchio2010}. In the following we adopt a fiducial value $s_1=2$.

As discussed in Section \ref{sec:motivation}, we make the further assumption that the momentum of the electrons becomes approximately aligned with the direction of the magnetic field when $\gamma\gtrsim\gamma_{\rm iso}$.\footnote{In principle, the dependence of the pitch angle on the Lorentz factor of the electrons at $\gamma\sim\gamma_{\rm iso}$ can be determined by the slope of the SED at frequencies $\nu\gtrsim\nu_{\rm s}$. This would require a detailed fit of the model to the SED of individual objects, which is out of the scope of the paper.} Since the synchrotron emission is suppressed when the pitch angle is close to zero, if $\gamma_{\rm iso}\lesssim\gamma_{\rm b}$ our model predicts the synchrotron radiation to peak at a frequency
\begin{equation}
\label{eq:nus}
\boxed{ \nu_{\rm s} = 3.7\times 10^6 \gamma_{\rm iso}^2 B \delta} \;,
\end{equation}
where $\delta$ is the Doppler factor, $\gamma_{\rm iso}$ is given by Eq. \eqref{eq:gamma_iso}, and $\nu_{\rm s}$ is measured in Hz. The peak luminosity can be written as
\begin{equation}
\label{eq:Ls1}
L_{\rm s} = V\delta^4 \int N\left(\gamma\right) P_{\rm s} \left(\gamma\right) {\rm d}\gamma \sim V\delta^4 N\left(\gamma_{\rm iso}\right) \gamma_{\rm iso} P_{\rm s} \left(\gamma_{\rm iso}\right)\;,
\end{equation}
where $V=4\pi R^3/3$ and
\begin{equation}
\label{eq:Ls2}
P_{\rm s} \left(\gamma_{\rm iso}\right) = \frac{4}{3}\sigma_{\rm T}c U_{\rm B} \gamma_{\rm iso}^2 \;,
\end{equation}
being $U_B=B^2/8\pi$ the magnetic energy density.\footnote{Note that we are assuming the Doppler amplification to be proportional to $\delta^4$, which is the appropriate case if the emitting region moves together with the fluid at the same velocity ${\bf v}$. If instead the emitting region is stationary and the fluid inside moves with uniform velocity ${\bf v}$ (i.e., it turns ``on and off'' as it enters and leaves the emitting region), the amplification would be proportional to $\delta^3/\Gamma_{\rm jet}$ \citep[e.g.][]{LindBlandford1985, Sikora1997}.} Combining Eqs. \eqref{eq:Ls1} and \eqref{eq:Ls2} we finally get
\begin{equation}
\label{eq:Ls}
\boxed{L_{\rm s} = \frac{2}{9}\sigma_{\rm T}c B^2 R^3 K \gamma_{\rm iso} \delta^4}\;,
\end{equation}
where we have used our fiducial $s_1=2$. Here $c$ is the speed of light and $\sigma_{\rm T}$ is the Thompson cross section.

The non-thermal electrons scatter the synchrotron photons to produce the IC peak of the SED. The resulting spectrum depends on the scattering regime of the photons at the synchrotron peak. These photons are scattered in the Thompson regime if
\begin{equation}
\label{eq:cond}
\gamma_{\rm b}h\nu_{\rm s}<\delta m_e c^2\;,
\end{equation}
and in the Klein-Nishina regime otherwise. Here $h$ is the Planck constant and $m_e$ is the electron mass. We discuss the two cases separately below.

\subsection{Thompson regime}
\label{sec:thompson}

The peak of the IC component is produced by the electrons at the break scattering the photons at the synchrotron peak. The peak frequency can be calculated as
\begin{equation}
\label{eq:nuc}
\nu_{\rm c} = \frac{4}{3}\gamma_{\rm b}^2 \nu_{\rm s}\;.
\end{equation}
The peak luminosity can be written as
\begin{equation}
\label{eq:Lc1}
L_{\rm c} = V\delta^4 \int N\left(\gamma\right) P_{\rm c} \left(\gamma\right) {\rm d}\gamma \sim V\delta^4 N\left(\gamma_{\rm b}\right) \gamma_{\rm b} P_{\rm c} \left(\gamma_{\rm b}\right)\;.
\end{equation}
Here
\begin{equation}
\label{eq:Lc2}
P_{\rm c}\left(\gamma_{\rm b}\right) = \frac{4}{3}\sigma_{\rm T}c U_\gamma \gamma_{\rm b}^2 \;,
\end{equation}
being $U_\gamma=L_{\rm s}/4\pi R^2 c \delta^4$ the radiation energy density of the synchrotron photons. Combining Eqs. \eqref{eq:Lc1} and \eqref{eq:Lc2} we finally get
\begin{equation}
\label{eq:Lc}
L_{\rm c} = \frac{4}{9}\sigma_{\rm T}RK \gamma_{\rm b} L_{\rm s}\;,
\end{equation}
where we have used our fiducial $s_1=2$.

In order to calculate the cooling time, it is important to realise that in our model the radiative losses of the electrons at the break are dominated by the IC. Hence, the cooling time is
\begin{equation}
\label{eq:tcool1}
t_{\rm cool} = \frac{\gamma_{\rm b} m_e c^2}{P_{\rm c}\left(\gamma_{\rm b}\right)} \;.
\end{equation}
The ratio between the cooling time $t_{\rm cool}$ and the dynamical time $t_{\rm dyn}=R/c$ can be presented as
\begin{equation}
\label{eq:tcool}
\frac{t_{\rm cool}}{t_{\rm dyn}} = \frac{3\pi m_e c^3 R \delta^4}{\sigma_{\rm T} L_{\rm s} \gamma_{\rm b}}\;.
\end{equation}

\subsection{Klein-Nishina regime}

In this case the peak of the IC component is produced by the electrons at the break scattering the photons whose energy equals $m_e c^2$ in the electron's frame. By construction, the frequency of these photons is below the synchrotron peak. The frequency of the IC peak is then
\begin{equation}
\label{eq:nucKN}
\nu_{\rm c} = \frac{4}{3}\frac{m_e c^2}{h} \gamma_{\rm b} \delta\;.
\end{equation}

The calculation of $L_{\rm c}$ and $t_{\rm cool}$ can be carried out as in the Thompson regime, with the only difference that only the photons with frequency smaller than $\delta m_e c^2/h\gamma_{\rm b}=3\nu_{\rm c}/4\gamma_{\rm b}^2$ contribute to the effective $U_\gamma$. Since this suppresses $P_{\rm c}$ by a factor of $\left(3\nu_{\rm c}/4\gamma_{\rm b}^2\nu_{\rm s}\right)^{1/2}$, one finds
\begin{equation}
L_{\rm c} = \frac{4}{9}\sigma_{\rm T}RK L_{\rm s} \left(\frac{3\nu_{\rm c}}{4\nu_{\rm s}}\right)^{1/2}
\end{equation}
and
\begin{equation}
\frac{t_{\rm cool}}{t_{\rm dyn}} = \frac{3\pi m_e c^3 R \delta^4}{\sigma_{\rm T} L_{\rm s}}  \left(\frac{4\nu_{\rm s}}{3\nu_{\rm c}}\right)^{1/2} \;,
\end{equation}
which are the analogous of Eqs. \eqref{eq:Lc} and \eqref{eq:tcool} respectively.

\subsection{Final remarks}

The requirement that the observed emission varies on time scales comparable with the dynamical time puts one more constraint to the model, namely
\begin{equation}
\label{eq:tvar}
\boxed{R=c\delta t_{\rm var}}\;.
\end{equation}
where $t_{\rm var}$ is the observed variability time scale. Substituting $\gamma_{\rm b}$ from Eq. \eqref{eq:nuc} into Eq. \eqref{eq:Lc}, one finds
\begin{equation}
\label{eq:tau}
\boxed{\sigma_{\rm T}RK = \frac{9L_{\rm c}}{4L_{\rm s}} \left(\frac{4\nu_{\rm s}}{3\nu_{\rm c}}\right)^{1/2}}\;.
\end{equation}
It is simple to realise that Eq. \eqref{eq:tau} is valid also in the Klein-Nishina regime. Note that Eqs. \eqref{eq:nus}, \eqref{eq:Ls}, \eqref{eq:tvar}, \eqref{eq:tau} do not contain $\gamma_{\rm b}$ any more, which allows one to determine $K$, $R$, $B$, $\delta$. Combining Eqs. \eqref{eq:nuc} and \eqref{eq:nucKN}, and taking into account the condition \eqref{eq:cond}, one eventually finds
\begin{equation}
\label{eq:gammab}
\boxed{\gamma_{\rm b}=\max\left[\left(\frac{3\nu_{\rm c}}{4\nu_{\rm s}}\right)^{1/2}, \left(\frac{3h\nu_{\rm c}}{4\delta m_e c^2}\right)\right]}\;,
\end{equation}
which proves that, once $\gamma_{\rm iso}$ is known, all the free parameters of the model are constrained by observations.

If the scattering occurs in the Thompson regime, one can calculate $t_{\rm cool}/t_{\rm dyn}$ isolating $\gamma_{\rm b}$ from Eq. \eqref{eq:nuc} and $R$ from Eq. \eqref{eq:tvar}, and substituting them into Eq. \eqref{eq:tcool}. This finally gives
\begin{equation}
\label{eq:delta}
\frac{t_{\rm cool}}{t_{\rm dyn}} = \frac{3\pi m_e c^4 t_{\rm var}}{\sigma_{\rm T} L_{\rm s}}  \left(\frac{4\nu_{\rm s}}{3\nu_{\rm c}}\right)^{1/2} \delta^5 \;,
\end{equation}
which is valid also in the Klein-Nishina regime.

\section{Comparison with an isotropic model}
\label{sec:comparison}

\begin{figure*}{\vspace{3mm}} 
\centering
\includegraphics[width=0.49\textwidth]{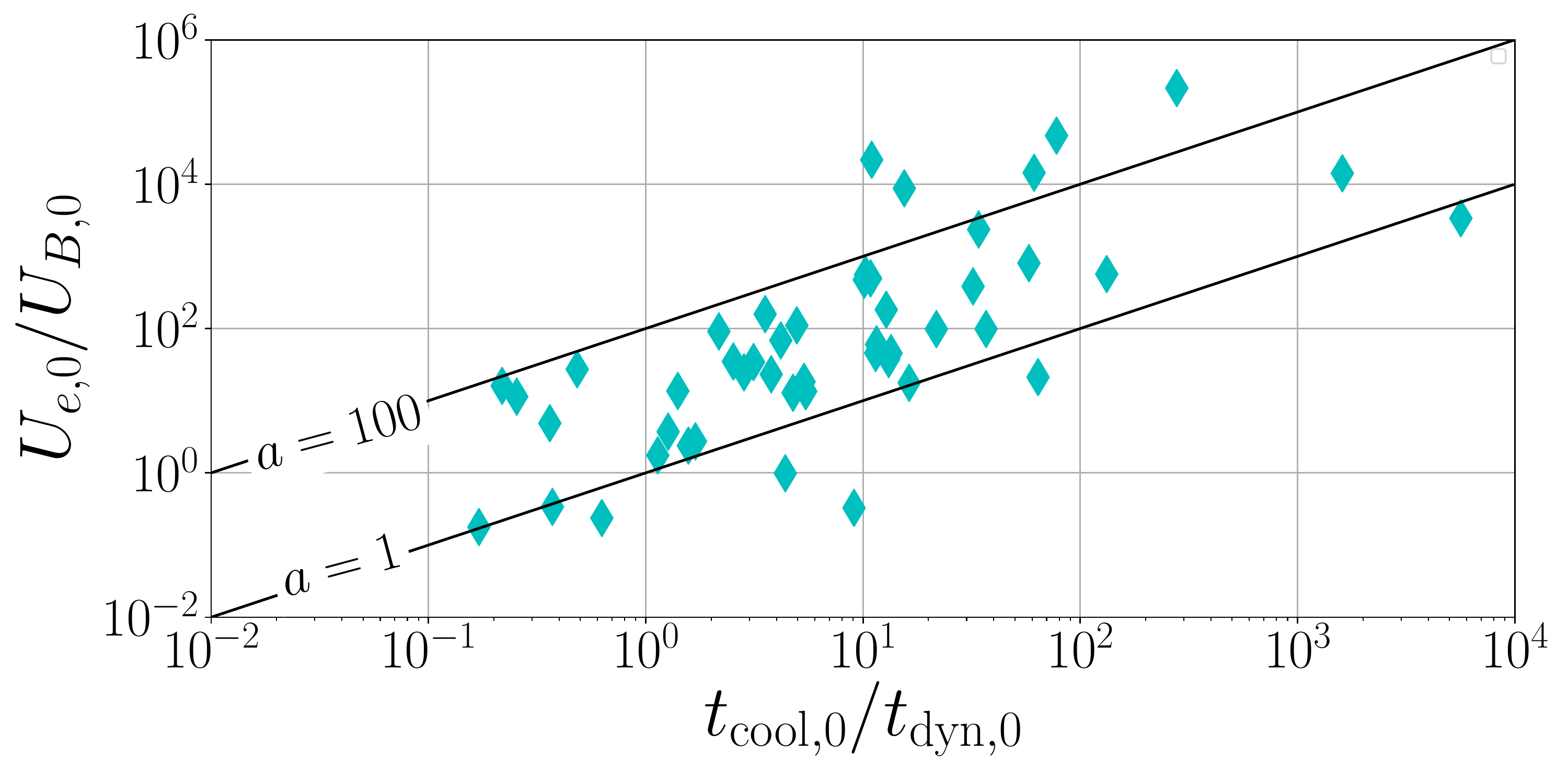}
\includegraphics[width=0.49\textwidth]{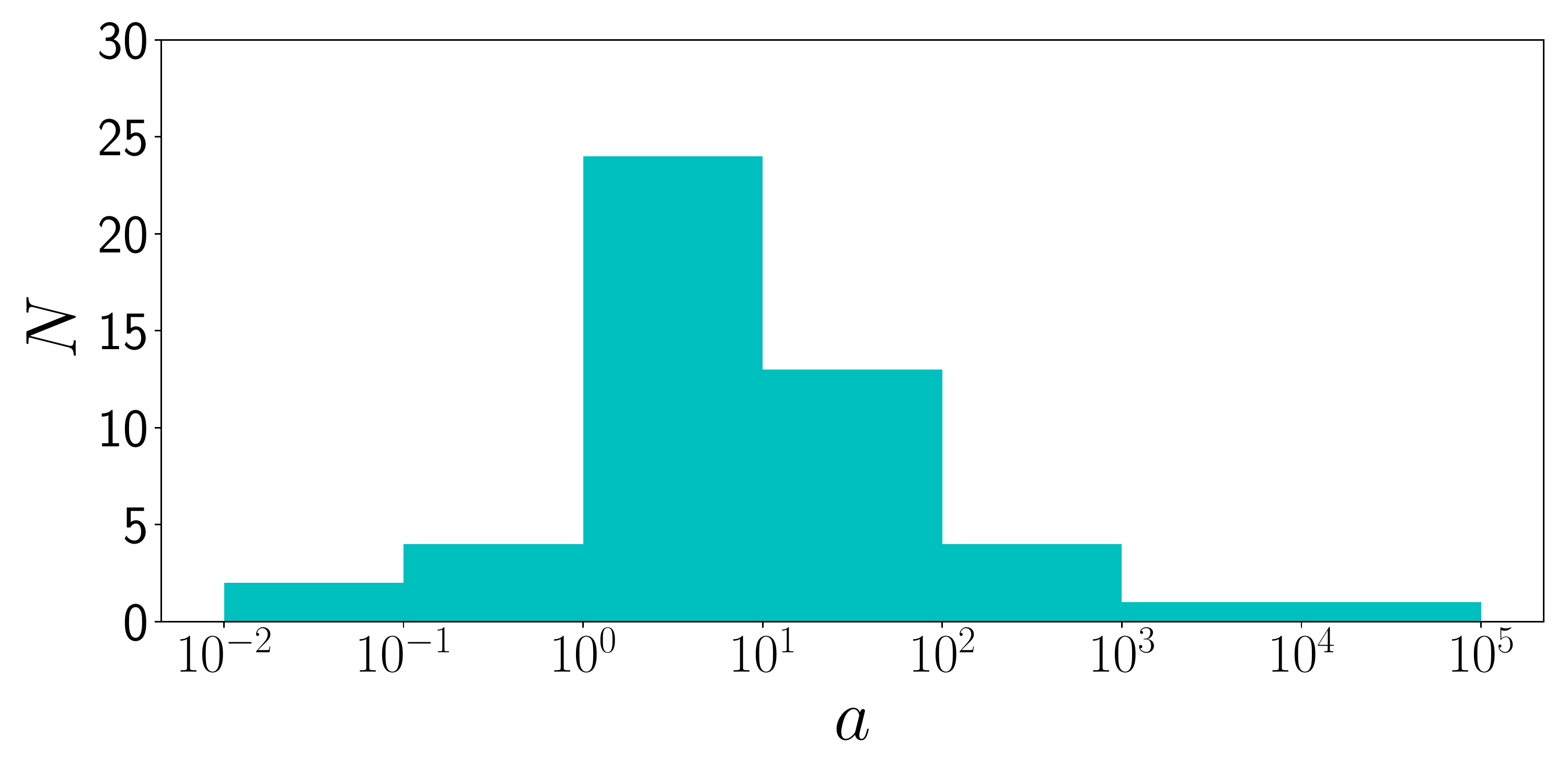}
\caption{Left panel: ratio of the energy carried by the electrons to the energy carried by the magnetic fields ($U_{e,0}/U_{B,0}$) versus ratio of the cooling to dynamical times ($t_{\rm cool,0}/t_{\rm dyn,0}$) in a one-zone SSC model that assumes an isotropic electron distribution. Each point represents a BL Lac in the sample of \citet{Tavecchio2010}. Right panel: distribution of the model-independent parameter $a\sim \left(U_e/U_B\right) \left(t_{\rm dyn}/t_{\rm cool}\right)^{19/15}$ for all the BL Lacs in the sample.}
\label{fig:prediction}
\end{figure*}

\subsection{Predictions of an isotropic model}

The usual one-zone SSC model has five free parameters, which we will name $K_0$, $\gamma_{\rm b0}$, $R_0$, $B_0$, $\delta_0$. Since the momentum distribution of the non-thermal electrons is assumed to be isotropic, both the synchrotron and the IC peaks of the SED are produced by the electrons at the break. The frequency and the luminosity of the synchrotron peak are given by
\begin{align}
\label{eq:nus0}
\nu_{\rm s} & = 3.7\times 10^6 \gamma_{\rm b0}^2 B_0 \delta_0\\
\label{eq:Ls0}
L_{\rm s} & = \frac{2}{9}\sigma_{\rm T}c B_0^2 R_0^3 K_0 \gamma_{\rm b0} \delta_0^4\;.
\end{align}
It is simple to realise that Eqs. \eqref{eq:tvar} and \eqref{eq:tau} are still valid. Hence
\begin{align}
\label{eq:tvar0}
R_0 & =c\delta_0 t_{\rm var}\\
\label{eq:tau0}
\sigma_{\rm T}R_0 K_0 & = \frac{9L_{\rm c}}{4L_{\rm s}} \left(\frac{4\nu_{\rm s}}{3\nu_{\rm c}}\right)^{1/2} \;.
\end{align}
The analogous of Eq. \eqref{eq:gammab} also holds, which finally constrains all the free parameters of the model from the observations.

In order to make a straightforward comparison with our model, here we define $t_{\rm cool,0}$ taking into account the IC cooling only. One finds
\begin{equation}
\label{eq:ratio}
\frac{t_{\rm cool,0}}{t_{\rm dyn,0}} = \frac{3\pi m_e c^4 t_{\rm var}}{\sigma_{\rm T} L_{\rm s}}  \left(\frac{4\nu_{\rm s}}{3\nu_{\rm c}}\right)^{1/2} \delta_0^5\;.
\end{equation}
Note that, since when the momentum distribution is isotropic the synchrotron cooling should be also taken into account, our $t_{\rm cool, 0}$ is longer than the true cooling time by a factor $1+P_{\rm s}/P_{\rm c}=1+L_{\rm s}/L_{\rm c}$. Since typically $L_{\rm s}$ and $L_{\rm c}$ are of the same order, which can be inferred directly from the SED of individual BL Lacs \citep[e.g.][]{Tavecchio2010}, this correction is a factor of a few.

\subsection{Relation between the physical parameters}

Since the observed quantities (namely $\nu_{\rm s}$, $\nu_{\rm c}$, $L_{\rm s}$, $L_{\rm c}$, $t_{\rm var}$) are model-independent by definition, one can find a relation between the physical parameters of our model and those of an isotropic model. Comparing Eq. \eqref{eq:nus} to Eq. \eqref{eq:nus0} it is simple to realise that
\begin{equation}
\label{eq:1}
\gamma_{\rm iso}^2 B \delta = \gamma_{\rm b0}^2 B_0 \delta_0\;.
\end{equation}
Comparing Eqs. \eqref{eq:Ls} and \eqref{eq:Ls0}, one finds
\begin{equation}
\label{eq:2}
B^2 R^3 K \gamma_{\rm iso} \delta^4 = B_0^2 R_0^3 K_0 \gamma_{\rm b0} \delta_0^4\;.
\end{equation}
Comparing Eqs. \eqref{eq:tvar} and \eqref{eq:tvar0}, one finds
\begin{equation}
\label{eq:3}
R/\delta = R_0/\delta_0\;.
\end{equation}
Finally, comparing Eqs. \eqref{eq:tau} and \eqref{eq:tau0}, one finds
\begin{equation}
\label{eq:4}
R K = R_0 K_0\;.
\end{equation}
One can solve Eqs. \eqref{eq:1}-\eqref{eq:4} in order to express the four parameters $K$, $R$, $B$, $\delta$ as a function of the others, which gives
\begin{align}
\label{eq:K_final}
K & =  \left(\frac{\gamma_{\rm b0}}{\gamma_{\rm iso}}\right)^{3/4} K_0\\
\label{eq:R_final}
R & =  \left(\frac{\gamma_{\rm iso}}{\gamma_{\rm b0}}\right)^{3/4} R_0\\
\label{eq:B_final}
B & = \left(\frac{\gamma_{\rm b0}}{\gamma_{\rm iso}}\right)^{11/4} B_0\\
\label{eq:delta_final}
\delta & = \left(\frac{\gamma_{\rm iso}}{\gamma_{\rm b0}}\right)^{3/4} \delta_0\;,
\end{align}
where $\gamma_{\rm iso}$ is given by Eq. \eqref{eq:gamma_iso}. These expressions give a simple correspondence between the parameters of the two models.

We are now in the position to evaluate how the two models differ in the predicted ratio of (i) the electron to the magnetic energy, and (ii) the cooling to dynamical times. Let $U_B$ and $U_e$ be the energy density of the magnetic fields and the kinetic energy density of the non-thermal electrons respectively. Since $U_B=B^2/8\pi$, we have $U_B=\left(B/B_0\right)^2 U_{B,0}$. One can calculate $U_e=\int \gamma m_ec^2 N\left(\gamma\right) {\rm d}\gamma \sim Km_e c^2 \log\left(\gamma_{\rm b}/\gamma_{\rm min}\right)$, where we have used the distribution \eqref{eq:N} with $s_1=2$. Neglecting the weak (logarithmic) dependence on $\gamma_{\rm min}$ and $\gamma_{\rm b}$, one sees that $U_e=\left(K/K_0\right)U_{e,0}$, from which it immediately follows that $U_e/U_B=\left(B_0/B\right)^2 \left(K/K_0\right) \left(U_{e,0}/U_{B,0}\right)$. Using Eqs. \eqref{eq:K_final} and \eqref{eq:B_final} we finally get
\begin{equation}
\label{eq:sigma}
\frac{U_e}{U_B} = \left(\frac{\gamma_{\rm iso}}{\gamma_{\rm b0}}\right)^{19/4} \frac{U_{e,0}}{U_{B,0}}\;.
\end{equation}
Comparing Eqs. \eqref{eq:delta} and \eqref{eq:ratio}, one sees that $t_{\rm cool}/\delta^5 t_{\rm dyn} = t_{\rm cool,0}/\delta_0^5 t_{\rm dyn,0}$, which using Eq. \eqref{eq:delta_final} gives
\begin{equation}
\label{eq:time}
\frac{t_{\rm cool}}{t_{\rm dyn}} = \left(\frac{\gamma_{\rm iso}}{\gamma_{\rm b0}}\right)^{15/4} \frac{t_{\rm cool,0}}{t_{\rm dyn,0}} \;.
\end{equation}
One sees that, in the case $\gamma_{\rm iso}\lesssim\gamma_{\rm b0}$, the ratio of both (i) the electron to the magnetic energy, and (ii) the cooling to the dynamical times predicted by our model can be significantly lower than those predicted by the usual isotropic model.

\section{Results}
\label{sec:results}

Our goal is showing that, assuming that the electron distribution becomes anisotropic at the highest energies, it is possible to have (i) an approximate equipartition between the energy carried by the non-thermal electrons and by the magnetic fields ($U_e\sim U_B$); (ii) the electrons at the break efficiently cooling in a dynamical time ($t_{\rm cool}\lesssim t_{\rm dyn}$). Combining Eqs. \eqref{eq:sigma} and \eqref{eq:time}, we see that
\begin{equation}
\label{eq:prediction}
\frac{U_e}{U_B} \left(\frac{t_{\rm dyn}}{t_{\rm cool}}\right)^{19/15} \sim \frac{U_{e,0}}{U_{B,0}} \left(\frac{t_{\rm dyn,0}}{t_{\rm cool,0}}\right)^{19/15}\;.
\end{equation}
In the following we define the parameter
\begin{equation}
a \sim \frac{U_e}{U_B} \left(\frac{t_{\rm dyn}}{t_{\rm cool}}\right)^{19/15} \;.
\end{equation}
Due to Eq. \eqref{eq:prediction}, such a parameter is model-independent (namely, $a\sim a_0$). If $U_e\sim U_B$ and $t_{\rm cool}\lesssim t_{\rm dyn}$ in the anisotropic model, one would expect $a$ to be distributed above a minimum value of order unity.

In order to calculate the parameter $a$ for individual BL Lacs, we use the results of \citet{Tavecchio2010}, who fitted the SED of a sample of BL Lacs using an isotropic model for the electron distribution. In the left panel of Figure \ref{fig:prediction} we plot $U_{e,0}/U_{B,0}$ versus $t_{\rm cool,0}/t_{\rm dyn,0}$ for all the BL Lacs in the sample of \citet{Tavecchio2010}.\footnote{In order to calculate $t_{\rm cool,0}$, (i) we calculate the cooling time due to synchrotron losses only, and (ii) we rescale it by a factor $L_{\rm s}/L_{\rm c}$. We infer the luminosities directly from the SED of individual objects.} One sees that, while both $U_{e,0}/U_{B,0}$ and $t_{\rm cool,0}/t_{\rm dyn,0}$ have a large scatter and most of the BL Lacs have $U_{e,0}/U_{B,0}\gtrsim 1$ and $t_{\rm cool,0}/t_{\rm dyn,0}\gtrsim 1$, these two quantities are correlated and the combination $a_0\sim \left(U_{e,0}/U_{B,0}\right) \left(t_{\rm dyn,0}/t_{\rm cool,0}\right)^{19/15}$ is slightly bigger than unity for the majority of the BL Lacs. In the right panel of Figure \ref{fig:prediction} we show the distribution of $a$ for all the BL Lacs in the sample. Approximately $50\%$ of the BL Lacs have $1\lesssim a\lesssim 10$ and $\sim 75\%$ of them have $1\lesssim a\lesssim 100$, while only $\sim 12\%$ of the objects have $a\lesssim 1$. As discussed above, this shows that it is possible to construct an anisotropic model for the electron momentum distribution such that $U_e\sim U_B$ and $t_{\rm cool}\lesssim t_{\rm dyn}$.

\subsection{Proof-of-concept: an anisotropic model with $\pmb{U_e\sim U_B}$}

\begin{figure}{\vspace{3mm}} 
\centering
\includegraphics[width=0.49\textwidth]{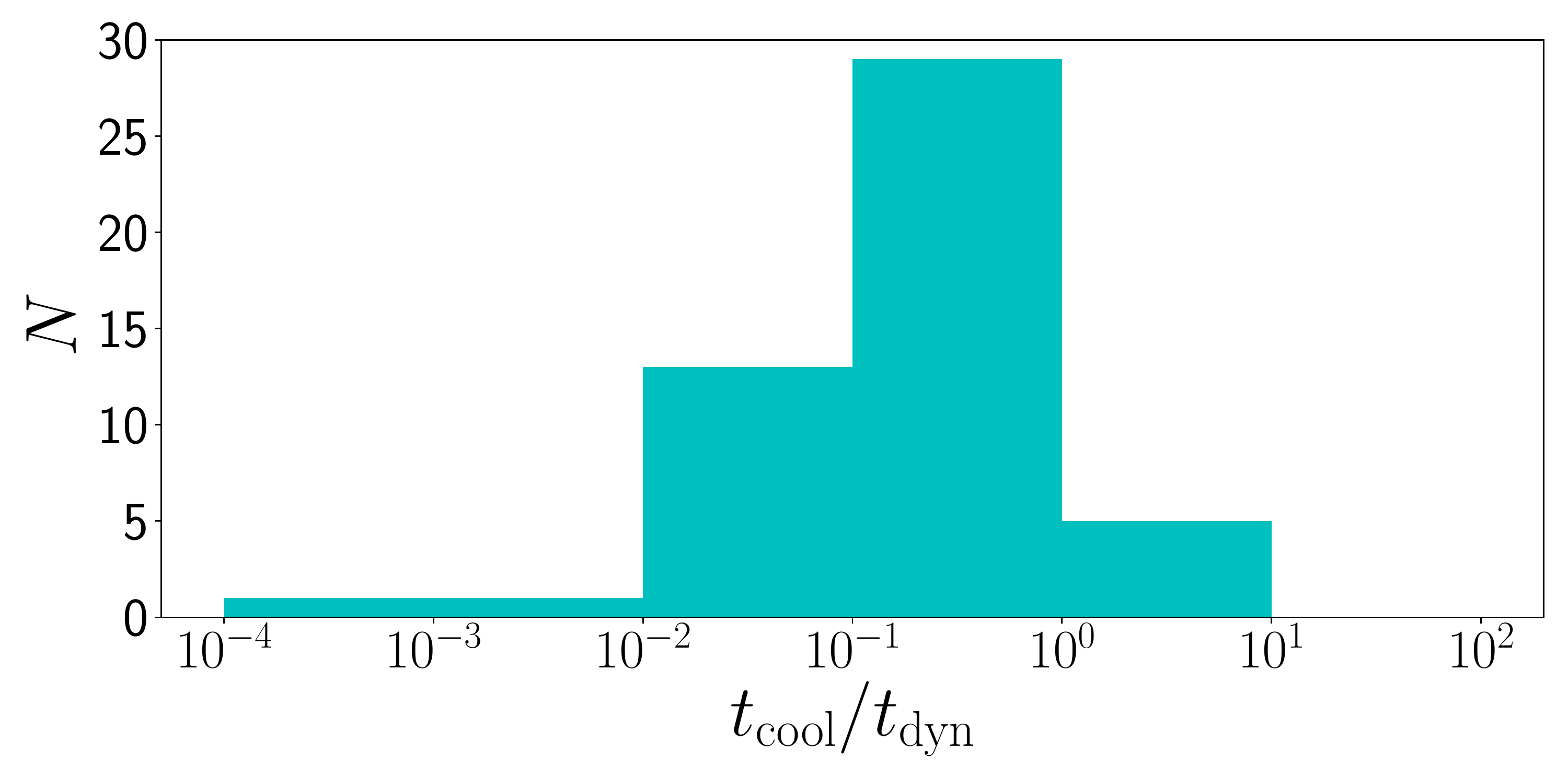}
\caption{Distribution of the ratio between the cooling to the dynamical times ($t_{\rm cool}/t_{\rm dyn}$) for all the BL Lacs in the sample. We assume an anisotropic model with energy equipartition between the non-thermal electrons and the magnetic fields ($U_e\sim U_B$).}
\label{fig:times}
\end{figure}

\begin{figure}{\vspace{3mm}} 
\centering
\includegraphics[width=0.49\textwidth]{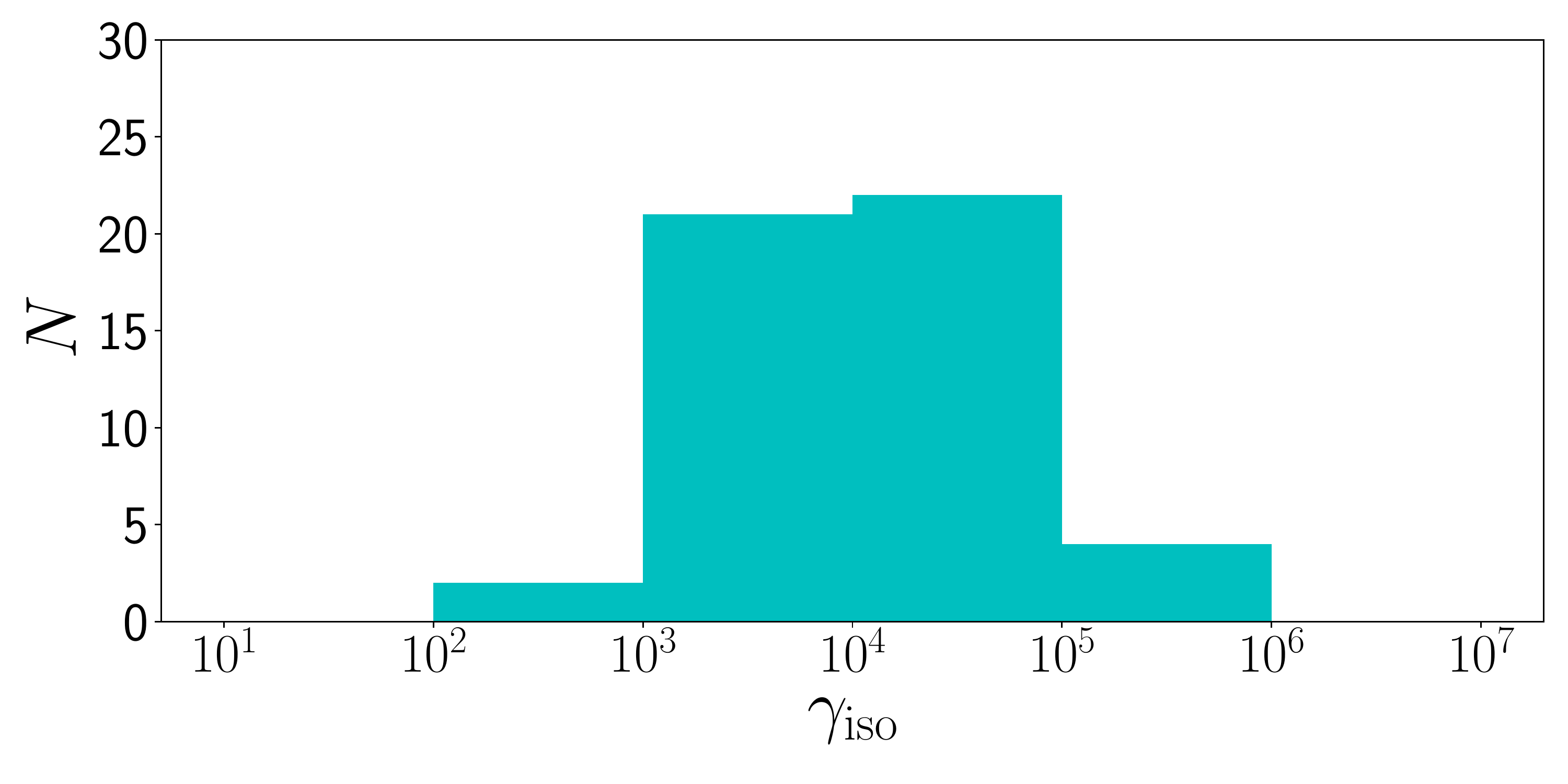}
\caption{Distribution of $\gamma_{\rm iso}$ for all the BL Lacs in the sample. We assume an anisotropic model with energy equipartition between the non-thermal electrons and the magnetic fields ($U_e\sim U_B$).}
\label{fig:gammaiso}
\end{figure}

\begin{figure*}{\vspace{3mm}} 
\centering
\includegraphics[width=0.49\textwidth]{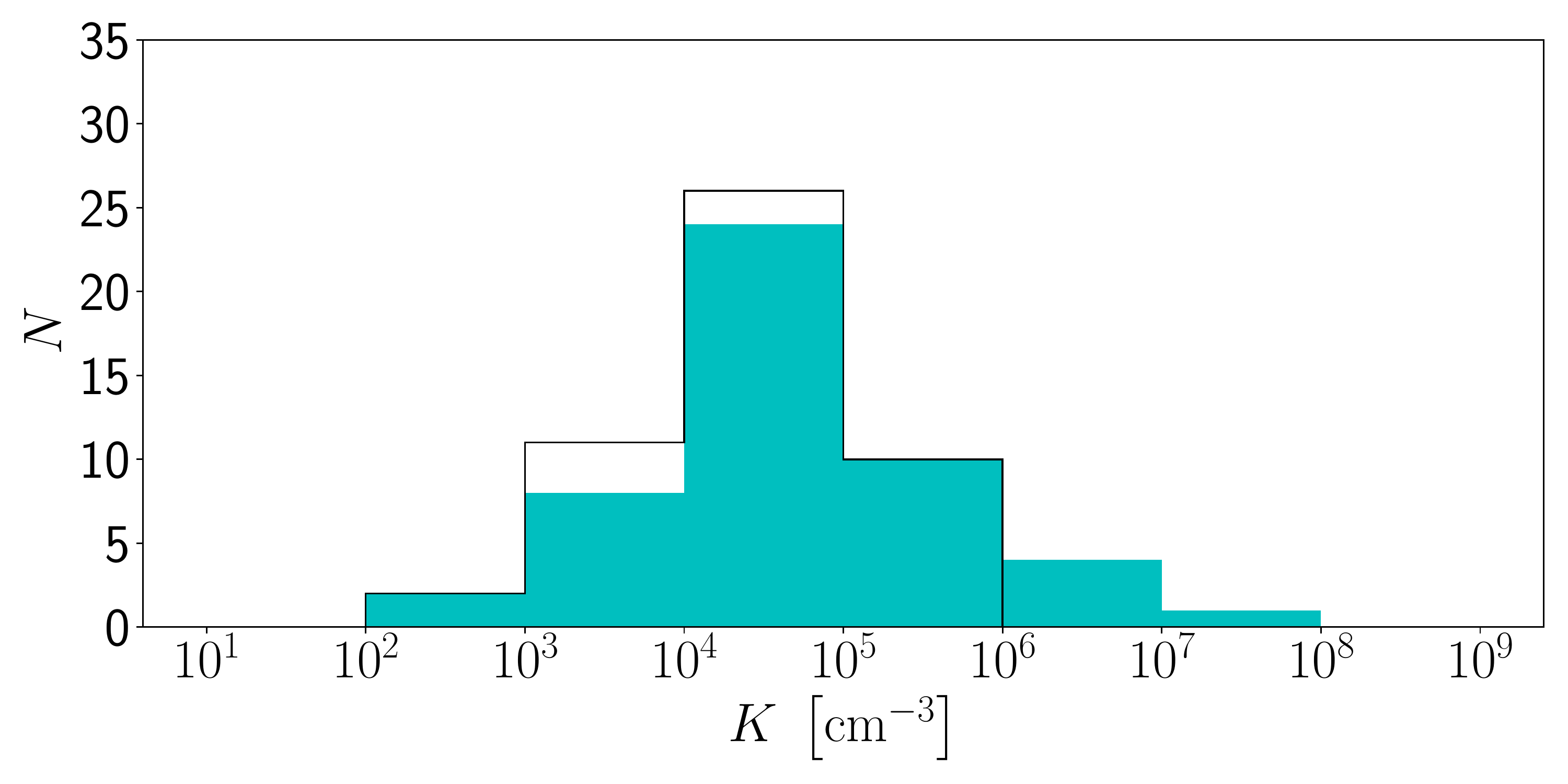}
\includegraphics[width=0.49\textwidth]{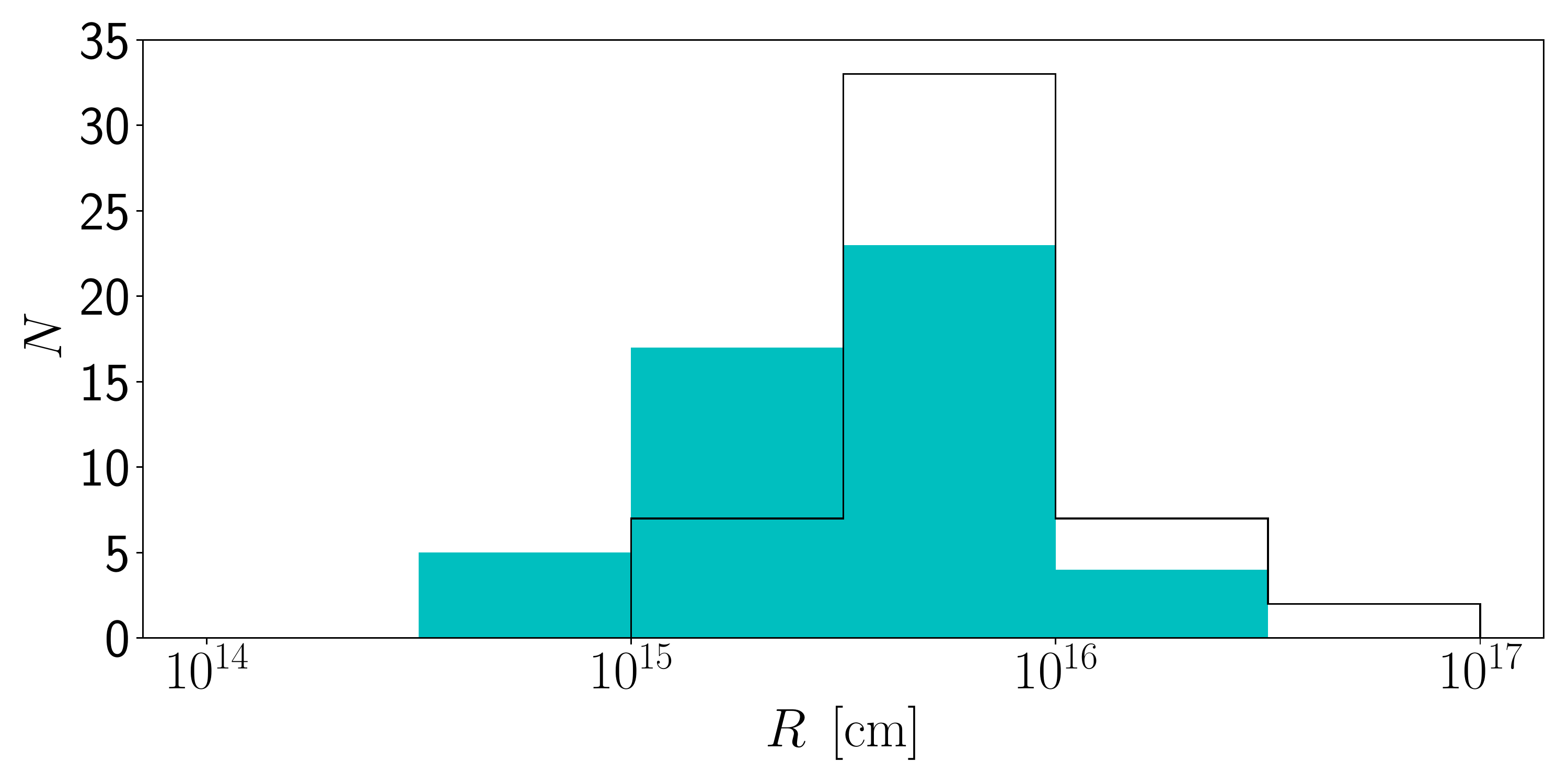}
\includegraphics[width=0.49\textwidth]{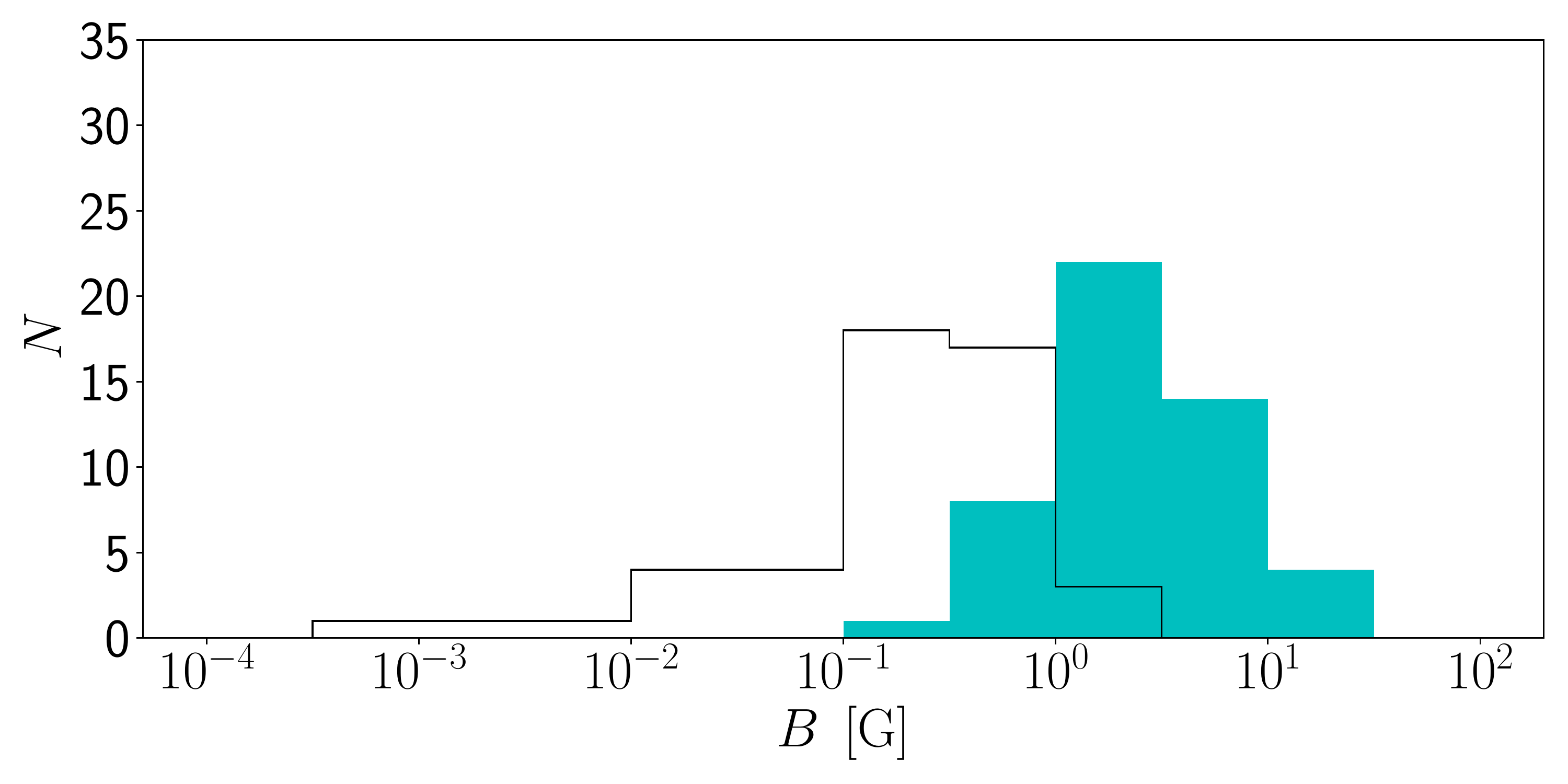}
\includegraphics[width=0.49\textwidth]{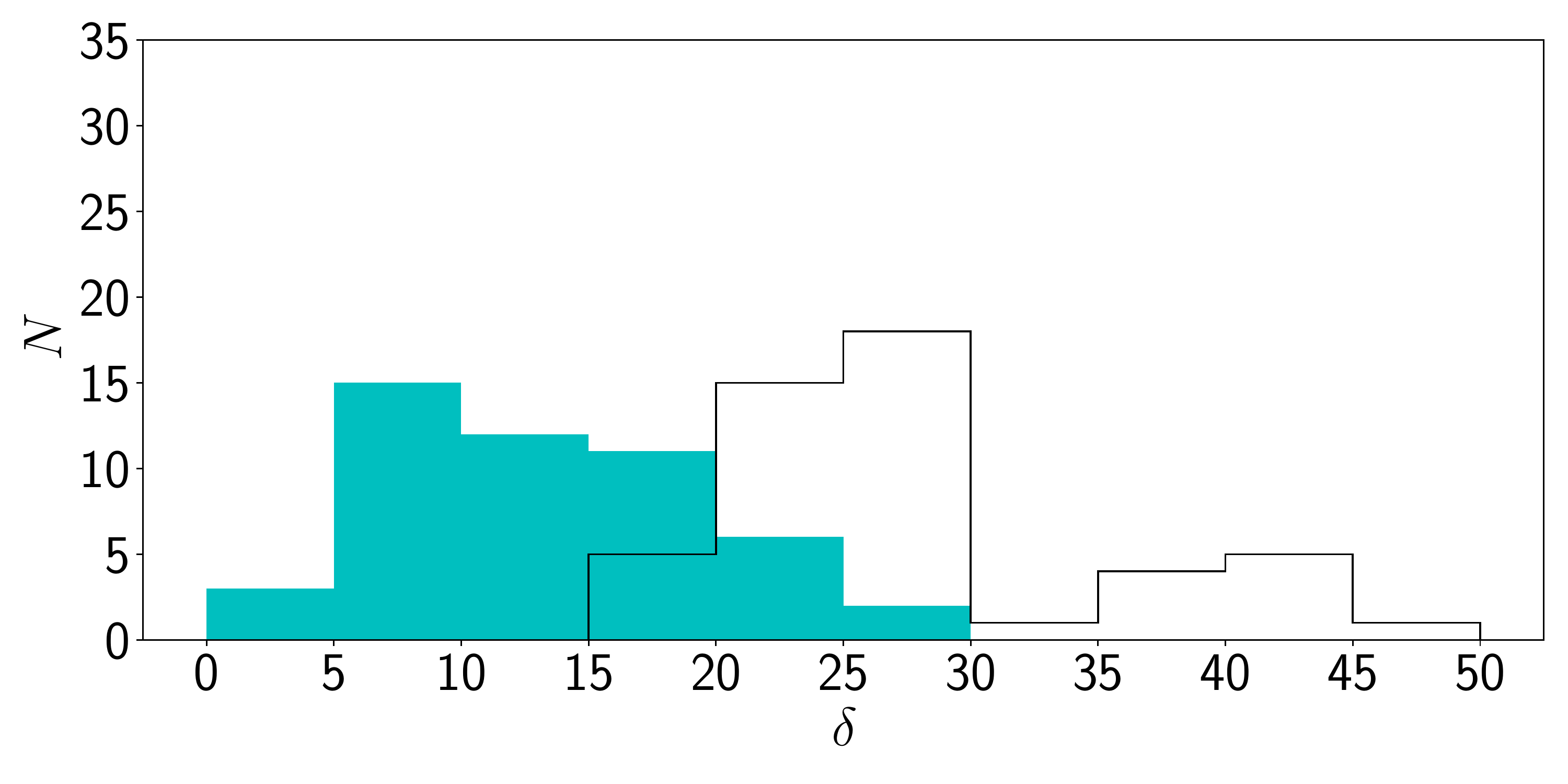}
\caption{Number density $K$ of the non-thermal electrons (top-left), size $R$ (top-right), magnetic field $B$ (bottom-left) and Doppler factor $\delta$ (bottom-right) of the dissipation region. The cyan histogram corresponds to an anisotropic model with energy equipartition between the non-thermal electrons and the magnetic fields ($U_e\sim U_B$). For comparison, with the thin line we show the distribution that is obtained assuming an isotropic electron distribution.}
\label{fig:parameters}
\end{figure*}

As an illustrative example, in the following we calculate all the parameters of our anisotropic model under the assumption that that $U_e\sim U_B$.\footnote{Five BL Lacs in the sample of \citet{Tavecchio2010} have $U_{e,0}\lesssim U_{B,0}$. For these objects we use the same best fit parameters of the isotropic model adopted by these authors.} In Figure \ref{fig:times} we show the distribution of $t_{\rm cool}/t_{\rm dyn}$, which we calculate using Eq. \eqref{eq:time} after finding $\gamma_{\rm iso}$ from Eq. \eqref{eq:sigma}. Approximately $60\%$ of the BL Lacs have $0.1\lesssim t_{\rm cool}/t_{\rm dyn}\lesssim 1$ and $\sim 85\%$ of them have $0.01\lesssim t_{\rm cool}/t_{\rm dyn}\lesssim 1$, while only $\sim 10\%$ of the objects have $t_{\rm cool}/t_{\rm dyn}\gtrsim 1$. This shows that an anisotropic model with $U_e\sim U_B$ naturally predicts the electrons at the break to cool efficiently in approximately a dynamical time. Moreover, since the ratio $t_{\rm cool}/t_{\rm dyn}$ is typically close to unity, one may speculate that the break in the energy distribution of the non-thermal electrons is separating the electrons whose cooling time is slower/faster than the dynamical time. Therefore the system is self-regulating: electrons are accelerated until they begin to loose the acquired energy. At any $\gamma_{\rm iso}$, the parameter $\gamma_{\rm iso}/\gamma_{\rm b}$, which determines both $U_e/U_B$ and $t_{\rm cool}/t_{\rm dyn}$ is self-adjusted.

In Figure \ref{fig:gammaiso} we show the distribution of $\gamma_{\rm iso}$, namely the Lorentz factor above which the electron distribution becomes elongated in the direction of the magnetic field. As discussed in Section \ref{sec:motivation}, we expect $\gamma_{\rm iso}$ to be somewhat larger than the proton to electron mass ratio, $m_p/m_e$. Indeed, only four BL Lacs in the sample require $\gamma_{\rm iso}\lesssim m_p/m_e$, while the large majority ($\sim 84\%$) of the objects have $m_p/m_e \lesssim\gamma_{\rm iso}\lesssim 10^5$, which is in reasonable agreement with our initial prediction (Eq. \ref{eq:gamma_iso}).

In Figure \ref{fig:parameters} we show the parameters $K$, $R$, $B$, $\delta$ predicted by our model, which we find using Eqs. \eqref{eq:K_final}-\eqref{eq:delta_final}. For comparison, with the thin line we show the distribution that is obtained assuming an isotropic electron distribution. The distribution of the number density of the non-thermal electrons and the distribution of the size of the dissipation region do not change significantly. The distribution of the magnetic field becomes narrower, and the typical field is significantly higher ($\sim 73\%$ of the BL Lacs have $1\;{\rm G}\lesssim B \lesssim 10\;{\rm G}$). Finally, the distribution of the bulk Doppler factor becomes monotonically decreasing, with most of the objects in the range $5\lesssim\delta\lesssim 25$. Interestingly, even if this was not guaranteed a priori, we find only three objects with $\delta\lesssim 5$ (two of them have $4.5\lesssim\delta\lesssim 5$ and only one has $\delta\sim 2$). This reassures us that our results do not systematically violate the lower limit on $\delta$ that is obtained requiring that the dissipation region is optically thin for pair production ($\gamma\gamma\to e^+ e^-$; e.g.\;\citealt[e.g.][]{DondiGhisellini1995}).\footnote{We have checked this constraint not to be violated for any object in the sample of \citet{Fan2014}, who calculated the lower limit on the Doppler factor for 457 blazars. Their sample includes $\sim 80\%$ of the objects in our sample.}

\section{Conclusions}
\label{sec:conclusions}

In this paper we have closely investigated one of the key assumptions that is usually adopted to interpret the SED of blazars, namely that the momentum distribution of the non-thermal electrons emitting the observed radiation is isotropic. We have found that this assumption may be oversimplified. Indeed, if the magnetic energy is dissipated via a turbulent MHD cascade, particles are primarily accelerated along the background magnetic field. In a highly magnetised plasma, the momentum of the lowest energy electrons may be isotropised by resonant wave-particle interactions. However, this mechanism is likely inefficient for the highest energy electrons, which may therefore retain a small pitch angle.

Motivated by the physics of energy dissipation in turbulent magnetised plasmas, we have presented a simple anisotropic model where the angular distribution of the electrons momenta depends on the single parameter $\gamma_{\rm iso}$: the electron momentum distribution is isotropic if the Lorentz factor is $\gamma\lesssim\gamma_{\rm iso}$, while the pitch angle becomes negligibly small when $\gamma\gtrsim\gamma_{\rm iso}$. The physical parameters of the dissipation region that are derived from the SED modelling are significantly affected by the anisotropy of the electron momentum distribution when $\gamma_{\rm iso}$ is below the spectral break of the distribution (namely, $\gamma_{\rm iso}\lesssim\gamma_{\rm b}$), as might be the case in a significant fraction of BL Lacs. The reason for such a difference with respect to the isotropic scenario is that, if $\gamma_{\rm iso}\lesssim\gamma_{\rm b}$, the synchrotron peak of the SED is produced by the electrons with $\gamma\sim\gamma_{\rm iso}$, while the IC peak is produced by the electrons with $\gamma\sim\gamma_{\rm b}$.

We have shown that, with a reasonable choice of the single parameter $\gamma_{\rm iso}$, it may be possible to construct a one-zone model reproducing the SED of BL Lacs such that  (i) the energy carried by the non-thermal electrons and by the magnetic fields are in an approximate equipartition ($U_e\sim U_B$); (ii) the non-thermal electrons efficiently cool in a dynamical time ($t_{\rm cool}\lesssim t_{\rm dyn}$). As discussed in the introduction, the fact that $U_e\sim U_B$ and $t_{\rm cool}\lesssim t_{\rm dyn}$ is in good agreement with a number of theoretical and observational constraints on AGN jets.\footnote{\citet{Nemmen2012} found a radiative efficiency of about 15\% for AGN jets. Taking into account that in Poynting dominated jets the fraction of energy going to heat could hardly exceed 50\% \citep{Peer2017}, that the electron spectrum is broad, and there are also protons, one concludes that the cooling time at the break, $t_{\rm cool}$, could not be significantly larger than $t_{\rm dyn}$.} Our results may therefore help to solve a controversy that was pointed out by \citet{TavecchioGhisellini2016}: indeed, modelling the BL Lac SED with a one-zone Self-Synchro-Compton model that assumes an isotropic momentum distribution for the non-thermal electrons typically gives $U_e\gg U_B$ and $t_{\rm cool}\gg t_{\rm dyn}$. Also note that, since our model predicts the ratio $t_{\rm cool}/t_{\rm dyn}$ to be typically close to unity for the electrons at the break of the energy distribution, one may speculate such a break to be associated with the Lorentz factor above which the cooling time becomes shorter than the dynamical time.

The dissipation of the magnetic energy through a turbulent MHD cascade may also explain the rapid variability that is observed in the spectra of blazars. Since the magnetic field in the emitting region is tangled, the radiation in the proper frame is isotropic when averaged over a suitably long time. However, as originally proposed by \citet{Thompson2006} in the context of GRBs, a fast variability on short time scales may be produced due to the fact that the radiation from a locally anisotropic electron distribution is strongly beamed. In blazars, the high energy variability of the spectrum is often explained by ``jet in a jet'' scenarios that may result from the magnetic reconnection process inside the jet \citep[see for example][]{Giannios2009, Giannios2010, Nalewajko2011}. We argue that the emission of highly beamed radiation may instead be the generic product of the energy dissipation in magnetically dominated jets.

Throughout this paper we have mostly been concerned about the statistical properties of BL Lacs. Nevertheless, our model can be used to fit the SED of individual objects. In particular, there are a few objects in the sample of \citet{Tavecchio2010} whose SED is difficult to model assuming an isotropic momentum distribution. As discussed by these authors, the reason is that the fit would require extremely large Doppler factors and small magnetic fields. It would be interesting to see if our model helps to improve the quality of the fit for these objects.

Finally, the fact that the highest energy electrons may retain a small pitch angle is based on a number of assumptions, namely (i) the pairs dominate the total number density ($n_e\gg n_p$), but the protons dominate the total mass density ($n_p m_p\gg n_e m_e$) of the jet, which is motivated by a number of independent arguments in the literature \citep[e.g.][]{SikoraMadejski2000, GhiselliniTavecchio2010}; (ii) the dissipation of the magnetic energy heats the particles in the direction of the background magnetic field, which is likely the case if the magnetic energy is brought down to the dissipation scale by a turbulent MHD cascade; (iii) the protons are heated less efficiently than the pairs, which is suggested by an analogy with the behaviour of non-relativistic turbulent plasmas. Future studies focusing on the dissipation of the magnetic energy via relativistic MHD turbulence may help to test the correctness of our assumptions (ii) and (iii).

We have not discussed the case of FSRQ yet. In these objects, the strong Compton dominance (which implies that $U_B\ll U_\gamma$) has led different authors to argue that the most promising explanation for the IC peak of the SED is the Comptonization of the radiation provided by a broad-line region or a dusty molecular torus \citep[see for example][]{Sikora2009, Ghisellini2010}. Though the presence of an external photon field makes the detailed modelling of the SED more uncertain than for BL Lacs, it has been suggested that in FSRQ jets (i) the amount of energy carried by the non-thermal electrons is comparable to that carried by the magnetic fields, namely $U_e\sim U_B$; (ii) the electrons at the break cool efficiently in a dynamical time, namely $t_{\rm cool}\sim t_{\rm dyn}$ \citep[e.g.][]{Ghisellini2010, GhiselliniTavecchio2015}, which would make the interpretation of the model's results less problematic than for BL Lacs. Our model may hardly affect these conclusions in a statistically significant number of FSRQ. The reason is that the typical break Lorentz factor in FSRQ is $\gamma_{\rm b}\sim 10^2$ \citep[e.g.][]{Ghisellini2010}, which is smaller than our expected $\gamma_{\rm iso}\gtrsim 10^3$ (see the discussion in Section \ref{sec:motivation} and in particular Eq. \ref{eq:gamma_iso}). Hence, in FSRQ the electrons at the break may become approximately isotropic, and thus produce both the synchrotron and the IC peaks of the SED.

\section*{Acknowledgements}

The authors acknowledge useful discussions with Amir Levinson. This research has received funding from the Israeli Science Foundation (grant 719/14) and from the German Israeli Foundation for Scientific Research and Development (grant I-1362-303.7/2016).

\def\aap{A\&A}\def\aj{AJ}\def\apj{ApJ}\def\apjl{ApJ}\def\mnras{MNRAS}
\def\araa{ARA\&A}\def\physrep{PhR}\def\sovast{Sov. Astron.}\def\nar{NewAR}
\def\aapr{Astronomy \& Astrophysics Review}\def\apjs{ApJS}\def\nat{Nature}\def\na{New Astron.}
\def\prd{Phys. Rev. D}\def\pre{Phys. Rev. E}\def\pasp{PASP}
\bibliographystyle{mn2e}
\bibliography{2d}

\appendix
\section{Conditions for the resonance instability of Alfv\'{e}n waves}
\label{sec:appendix}

In Section \ref{sec:motivation} we have studied the stability of a highly magnetised plasma where the momenta of all the particles are aligned with the background magnetic field. Right circularly polarised Alfv\'{e}n waves are damped due to the absorption by the resonant electrons, while they are emitted by the resonant protons and positrons. Hence, the system is unstable if $n_{p{\rm,res}} + n_{e^+{\rm,res}} \gtrsim n_{e^-{\rm,res}}$. One can calculate the damping time scale from
\begin{equation}
\frac{\Delta p_{\rm wave}}{\Delta V\Delta t} \sim - \frac{\Delta p_{e^-{\rm ,res}}}{\Delta V \Delta t}\;,
\end{equation}
where
\begin{equation}
\frac{\Delta p_{\rm wave}}{\Delta V\Delta t} \sim -\frac{1}{t_{\rm damp}} \frac{\left(\delta B\right)^2}{8\pi v_{\rm A}}\;.
\end{equation}
Using Eq. \eqref{eq:absorption_p}, this finally gives
\begin{equation}
\label{eq:tdamp_e}
t_{\rm damp} \sim \frac{B}{8\pi v_{\rm A}e n_{e^-{\rm,res}} }\;.
\end{equation}
In a similar way, one can calculate the growth rate from
\begin{equation}
\frac{\Delta p_{\rm wave}}{\Delta V\Delta t} \sim \frac{\Delta p_{e^+{\rm ,res}}}{\Delta V \Delta t} + \frac{\Delta p_{p{\rm ,res}}}{\Delta V \Delta t}\;,
\end{equation}
where
\begin{equation}
\frac{\Delta p_{\rm wave}}{\Delta V\Delta t} \sim \frac{1}{t_{\rm growth}} \frac{\left(\delta B\right)^2}{8\pi v_{\rm A}}\;.
\end{equation}
Using Eqs. \eqref{eq:emission_p} and \eqref{eq:pwave1}, this finally gives
\begin{equation}
t_{\rm growth} \sim \frac{B}{8\pi v_{\rm A}e \left(n_{p{\rm,res}} + n_{e^+{\rm,res}}\right)}\;.
\end{equation}
One immediately sees that the condition $n_{p{\rm,res}} + n_{e^+{\rm,res}} \gtrsim n_{e^-{\rm,res}}$ is equivalent to $t_{\rm growth}\lesssim t_{\rm damp}$.

The instability is effective if $t_{\rm growth}$ and $t_{\rm damp}$ are short with respect to the other relevant time scales of the system. Since the instability develops once $t_{\rm growth}\lesssim t_{\rm damp}$, it is sufficient to check $t_{\rm damp}$ to be short. The wave packet considered above may suffer from the additional damping by the wave-wave interaction with the packets from the turbulent MHD cascade. Following \citet{FarmerGoldreich2004}, we estimate the time scale for turbulent damping as
\begin{equation}
t_{\rm turb}\sim\frac{\sqrt{R\lambda_\parallel}}{v_{\rm A}}\;,
\end{equation}
which is much shorter than the dynamical time in the relevant case $\lambda_\parallel\ll R$. Using Eq. \eqref{eq:nres_01} to calculate $n_{e^-{\rm ,res}}$, the ratio $t_{\rm damp}/t_{\rm turb}$ may be expressed as
\begin{equation}
\frac{t_{\rm damp}}{t_{\rm turb}} \sim \frac{B^2}{8\pi n_e m_e c^2} \sqrt{\frac{\lambda_\parallel}{R}}\;.
\end{equation}
If the electron and the magnetic energy density are in an approximate equipartition, one sees that $t_{\rm damp}/t_{\rm turb}\sim\sqrt{\lambda_\parallel/R}\ll 1$. Hence, $t_{\rm damp}$ is the relevant time scale of the system, being it of the order of the Larmor time of the resonating electrons.

Finally, note that the particle pitch angle may change due to synchrotron emission, which damps the perpendicular particle motion, and due to IC emission (since the photons are typically scattered exactly in the direction of motion, we argue that IC emission is quite inefficient to produce a diffusion in the pitch angle). The important point is that in our model the cooling time due to both these processes is comparable to the dynamical time (see Figure \ref{fig:times}). Since we have shown the instability discussed in Section \ref{sec:motivation} to operate on much shorter time scales, we expect that synchrotron and IC processes hardly affect the pitch angle distribution.

\end{document}